# Phase separation, effects of magnetic field and high pressure on charge ordering in γ-$Na_{0.5}CoO_2$

H.X. Yang*, Y.G. Shi, C.J. Nie, D. Wu, L.X. Yang, C. Dong, H.C. Yu, H.R. Zhang, C.Q. Jin, and J.Q. Li

Beijing National Laboratory for Condensed Matter Physics, Institute of Physics, Chinese Academy of Sciences, Beijing 100080, China

## Abstract

Transmission electron microscopy observations reveal the presence of complex superstructures and remarkable phase separation in association with Na-ordering phenomenon in γ-$Na_{0.5}CoO_2$. Resistivity and magnetization measurements indicate that three phase transitions at the temperatures of 25 K, 53 K and 90 K respectively appear commonly in γ-$Na_{0.5}CoO_2$ samples. Under a high pressure up to 10 Kbar, the low-temperature transport properties show certain changes below the charge order transition; under an applied magnetic field of 7 Tesla, phase transitions at around 25 K and 53 K, proposed fundamentally in connection with alternations of magnetic structure and charge ordering maintain almost unchanged.

Author to whom correspondence should be addressed: hxyang@blem.ac.cn

# 1. Introduction

The $Na_xCoO_2$ system has attracted much attention in past several years due to its large thermoelectric power coexisting with low electric resistivity [1-4]. The discovery of superconductivity in water-intercalated $Na_xCoO_2$ immediately spurred tremendous round of intense interest in this system [5-6]. A phase diagram of non-hydrated $Na_xCoO_2$ for $0.3 < x < 0.75$ has recently been reported, it shows two distinct metallic states separated by an insulating state that is stabilized at $x = 0.5$ by charge ordering [7]. The charge ordering phenomena has been suggested to be in connection with a major instability in the narrow conduction band of $CoO_2$ layer [8]. At low temperatures, there are three phase transitions (at around 87 K, 53 K, and 20 K respectively) have been well identified by the measurements of resistivity, thermal transport and magnetization [7]. Previous studies suggest that the 87 K transition arises partially from a structural change; the transitions at 53 K and 20 K are likely responsible for charge ordering associated with a magnetic ordering at 53 K and then a spin reorientation at 20 K [9]. Microstructure analysis by means of electron diffraction reveals the presence of an orthorhombic superstructure phase in $Na_{0.5}CoO_2$ attributed to Na ordering [10]. Neutron powder diffraction measurements further demonstrated an ordering of the Na ions into zigzag chains along one crystallographic direction, which decorates the chains of Co ions with different amounts of charges [9].

Previously, it is well demonstrated that the antiferromagnetic and charge order transitions can be evidently alternated by either applied magnetic fields or high pressures, as typically reported in high $T_c$ superconductors and colossal magnetoresistance Mn oxides [11-12]. In this paper, we report on the microstructure properties of the $\gamma$-$Na_{0.5}CoO_2$ materials in connection with Na ordering and modifications of low temperature phase transitions under the applied magnetic fields and high pressures.

## 2. Experimental

Ceramic pellets and single crystal $Na_{0.5}CoO_2$ samples were used in the present experiments. The ceramic $Na_{0.75}CoO_2$ was synthesized following a procedure as described in Ref. 13. High-quality $Na_{0.85}CoO_2$ single crystals were grown using a traveling-solvent floating zone method. $Na_{0.5}CoO_2$ compounds were prepared by sodium deintercalation of $Na_{0.85}CoO_2$ and $Na_{0.75}CoO_2$ compounds following Ref. 14. Specimens for transmission-electron microscopy (TEM) observations are prepared simply by crushing the bulk material into fine fragments, which were then supported by a copper grid coated with thin carbon film. The TEM investigations were performed on an H-9000NA TEM operating at the voltage of 300kV. The sodium content of all samples was determined by ICP method. Powder XRD was performed from room temperature down to 9 K by employing a RIGAKU x-ray diffractometer.

## 3. Results and discussion

### 3.1. TEM observations

TEM observations of the structural features of $Na_{0.5}CoO_2$ have been performed in both the single crystalline and polycrystalline samples from room temperature down to 100 K. In addition to the superstructures as observed in previous publications [10], evident phase segregation shown up as either regular or irregular structural lamella commonly appear in the $Na_{0.5}CoO_2$ materials. Fig. 1 (a)-(c) show the dark field TEM image, and electron diffraction patterns obtained from the domain A and domain B, respectively. This TEM image is obtained by using a satellite spot following with the (110) main reflection. The micrometer-scale regions with relatively bright uniform contrast are associated with superstructure phase. Stripe like lamella stretching along the <110> direction can be clearly recognized. Regions with darker contrast are in association with the conventional hexagonal structure and demonstrated by the diffraction pattern of Fig. 1 (b), in which no clear superstructure spots are visible. Fig. 1 (c) represents a typical diffraction pattern for the superstructure phase commonly appears in the x = 0.5 materials. The remarkable structural property for this superstructure phase is the presence of the incommensurability along the <110> direction. This feature is demonstrated with notable spot splits as indicated by arrows. Fig. 1 (d) shows a micro-photometric density curve clearly illustrating the spot splitting along the <110> direction. The presence of two peaks at the <110>/2 positions is demonstrated. As a matter of fact, all superstructure spots in present pattern can be well indexed by a structural modulation with a modulation wave vector $q = <1/4, 1/4, 0> + \delta$. Where the incommensurate parameter $\delta$, changing slightly from area to area, depends sensitively on temperature and electron beam illumination as found in our experiments. All superstructure spots in Fig. 1 (c) are at the systematic position of (LK0) + nq with L, K, n are integers. The superstructure spots in present case are generally very sharp and strong, directly indicating the long-coherent nature (> 100 nm) of the Na ordered state and presence of evident resultant structural distortion. Low-temperature TEM observations reveal that this incommensurate modulated structure changes to a well-defined orthorhombic structure below 140 K as illustrated in the electron diffraction pattern of Fig. 1 (e). This superstructure can be indexed on to an orthorhombic cell with lattice parameter of $a = \sqrt{3}a_{hex}$, $b = 2a_{hex}$, $c = c_{hex}$.

Following with the ways used previously for explaining the numerous superstructures appear in the $Na_xCoO_2$ system, the structural model with $Na_1$ - $Na_2$ chain order can be used in principle to explain our experimental results [9, 10]. Microstructure investigation indicates that the structural inhomogeneity also appear at the low temperature phase. Fig. 1 (f) shows an electron diffraction pattern demonstrating the presence of well-defined twining structure frequently appear at low temperature.

In both polycrystalline and single crystalline samples, we can also see some other kind of superstructures that in general are shown as weak superstructure spots or diffused reflection streaks in the electron diffraction patterns. For instance, Fig. 2 (a) shows a [001] zone axis electron diffraction pattern with weak diffused diffractions. These defused diffraction centered roughly at the systematic {4/5, 4/5, 0} positions and spread to three distinctive directions. TEM observations indicate that crystals with this kind of structure also show up clear contrast inhomogeneity possibly resulting from phase separation. TEM observations of the microstructure properties corresponding with this kind of superstructure were performed in the certain regions at room temperature. Fig. 2 (b) shows a dark-field TEM image of a thin area by using the (020) main spot. The submicrometer-scale bright region with relatively uniform contrast has the conventional hexagonal structure as further conformed by nano-diffraction technique. The small dark speckles are believed to be corresponding to Na-ordered areas as small as a few nanometers in size.   Actually, our systematic TEM experimental investigations on the $Na_xCoO_2$ samples with x in the large range of 0.3 to 0.85 reveal that multi-modulated structures in association with complex contrast, as unveiled in dark-field TEM images, common appear in the this layered system with high $Na^+$ mobility. Hence, complicated

Na-ordered state and phase segregations, as a noteworthy structure features, exist commonly in $Na_xCoO_2$ materials. These features could yield salient influence on the physical properties observed in present system, such as superconductivity, charge ordering transitions and thermoelectric power.

*3.2. Resistivity and magnetization measurements*

Charge ordering proposed appearing below 100 K in the x = 0.5 sample has been first investigated in several samples (both polycrystalline and single crystals) by measurements of resistivity and magnetization. Three phase transitions as reported in previous literatures are commonly recognizable but with notable difference. The most evident anomaly shown up as a sharp upturn in resistivity and a downturn in magnetization occurs at the temperature of around 53 K in all samples examined in our study. This fact suggests that the striking alternation in connection with CO and magnetic structure definitely appears at this critical point. The transition at lower temperature of about 25 K can be unambiguously recognized in resistivity but occasionally appears in magnetization. Previous studies suggest that this transition is association with ordering of two distinctive types of magnetic atoms. The transition at 90 K, proposed in connection with certain kind of structural changes, has limited effects on both resistivity and magnetization. Fig. 3 shows the resistivity and magnetic susceptibility (measured by zero-field cooling (ZFC) DC in a field of 20 Oe) as function of temperature, characterizing of the low temperature phase transitions in a polycrystalline $Na_{0.5}CoO_2$ sample. The evident anomaly of charge ordering in resistivity occurs at the temperature of around 53 K, a notable saturation just below 25 K, following by another increase of the slope, which suggests another phase transition at this temperature. Moreover, the magnetic susceptibility shown in this figure demonstrates three clear transitions at 90 K, 53 K and 25 K, respectively.

*3.3. X-ray powder diffraction analysis*

In order to understand the structure evolution along with these low-temperature phase transitions, we have performed a series of XRD measurements from 300 K down to 10 K, the results suggest that the structure change along with charge order transitions in present system is very small, no evident discontinuous changes in either lattice parameters or crystal symmetry at the critical temperatures are detected. This result is in sharp contrast with the data obtained for $La_{0.5}Ca_{0.5}MnO_3$ system in which a clear structural change (first order phase transition) commonly occurs in associated with charge ordering transitions [15]. Although the resolution of XRD is not high enough to elucidate the fine details of the structure evolution alone temperature, further examination still provides some useful information. Table 1 shows the structural Rietveld refinements using the conventional $P6_3/mmc$ symmetry. The c axis parameter almost keeps constant below the transition ($T_{co}$ ~ 50 K) and increase progressively with temperature above $T_{co}$. On the other hand, the *a*-axis parameter remains constant in the whole temperature range from 4 K to 300 K.

*3.4. Applied magnetic fields or high pressures effect*

Fig. 4 (a) shows the temperature dependence of resistivity of $Na_{0.5}CoO_2$ under different pressures ranging from 0 to 10 Kbar. These experimental results indicate that, under the high pressures, the resistivity is hardly affected above the charge order transition and decreases clearly below charge ordering temperature. Moreover, when the pressure is larger than 6 Kbar, remarkable resistivity reductions appear along with the increase of pressure below 50 K. For facilitating the comparison and for the sake of clarify Fig. 4 (b) shows the result of the derivatives of the resistivities obtained at the ambient pressure and a high pressure of 8 Kbar. Pressure usually tends to increase the electron orbital overlaps, which leads to the enhancement of the conductivity. This is in particularly obvious at the charge ordering state which is initially of a very high resistance due to the static charge pinning. It can be apparently recognized that charge order transition slightly shifts to higher temperature at both critical points of about T ~ 53 K and 25 K. These results demonstrate that the high pressure could result in certain changes on the electronic structure of low-temperature insulator $Na_{0.5}CoO_2$, for which the charge-ordering gap has opened fully [14].

In the studies of manganites, it has been well evident that the charge order transition can be much reduced by applying an external field, since the field forcedly aligns the local spins and hence the double-exchange carriers gain mobility due to the reduced spin scattering. This phenomenon is typically seen in temperature dependence of resistivity for the $La(Sr)MnO_3$ and $Pr(Sr)MnO_3$ materials [16]. In our recent investigations, the magnetic field effects on the low temperature phase transitions have been examined for $Na_{0.5}CoO_2$ materials. Fig. 5 illuminates the temperature dependence of resistivity (R-T curve) for $Na_{0.5}CoO_2$ in the absence of a magnetic field and under a magnetic field up to 7 Tesla. The results demonstrate that the transition upturns remains almost the same under 7 Tesla.

## 4. Conclusions

In summary, we have performed an extensively investigation on magnetic and high pressure effects on the charge order phase transitions in $Na_{0.5}CoO_2$. The results demonstrate that the critical temperatures slightly shift to higher temperature along with the increase of pressure. Under an applied magnetic field of 7 Tesla, the transition upturns in resistivity remains almost unchanged. Transmission electron microscopy investigations on the $\gamma$-$Na_{0.5}CoO_2$ materials reveal the presence of a rich variety of superstructures arising from Na ordering. Phase segregation shown up as complex micro domains in TEM dark field images commonly appears in this layered material, this structural feature essentially results from chemical inhomogeneity in connection with Na distribution and ordering.

## Acknowledgments

We would like to thank X.Z. Liu for his assistance in measuring some physical properties and professor N.L. Wang and J.L Luo of Institute of Physics, Chinese Academy of Sciences for helpful discussion. The work reported here was supported by the "Outstanding Youth Fund" organized by National Natural Foundation of China.

# Figure and Table Captions

Table 1   Lattice parameters of $Na_{0.5}CoO_2$ from 300 K down to 9 K

Figure 1   (a) The dark field TEM image showing the coexistence of two domains. (b) Diffraction pattern of domain A without superstructure spots. (c) Diffraction pattern of domain B, representing a typical diffraction pattern for the incommensurate structural modulation in the x = 0.5 material. (d) A micro-photometric density curve for diffraction pattern of domain B, clearly illustrating the spot splitting at along the <110>. (e) Diffraction pattern from the low temperature orthorhombic phase. (f) Diffraction pattern illustrating the presence of the low-temperature twining structure.

Figure 2   (a) the [001] zone axis electron diffraction pattern with weak diffused diffractions. (b) A dark-field TEM image of a thin area by using the (020) main spot.

Figure 3   The magnetic resistivity characterization of the phase transitions in polycrystalline $Na_{0.5}CoO_2$ below 100 K

Figure 4   (a) Temperature dependence of the resistivity for the pressure between 0 and 10 Kbar. (b) Typical curves after taking the derivative of Fig. 4a.

Figure 5   (a) Temperature dependence of resistivity for the magnetic fields of 0 and up to 7 Tesla. (b) Curves after taking the derivative of Fig. 5a, illustrating the evident changes of critical temperatures.

**Table 1**

| T(K) | a (Å) | c (Å) | c/a | V (Å$^3$) |
|------|-------|-------|-----|-----------|
| 9    | 2.82119 | 11.07529 | 3.92575 | 229.01941 |
| 35   | 2.81993 | 11.07643 | 3.92791 | 228.83844 |
| 65   | 2.82132 | 11.07857 | 3.92673 | 229.10835 |
| 105  | 2.82027 | 11.08098 | 3.92905 | 228.98765 |
| 150  | 2.82152 | 11.09649 | 3.93281 | 229.51148 |
| 200  | 2.82250 | 11.11052 | 3.93641 | 229.96133 |
| 250  | 2.81960 | 11.12475 | 3.94551 | 229.78294 |
| 300  | 2.82132 | 11.14321 | 3.94964 | 230.44513 |

**Figure 1**

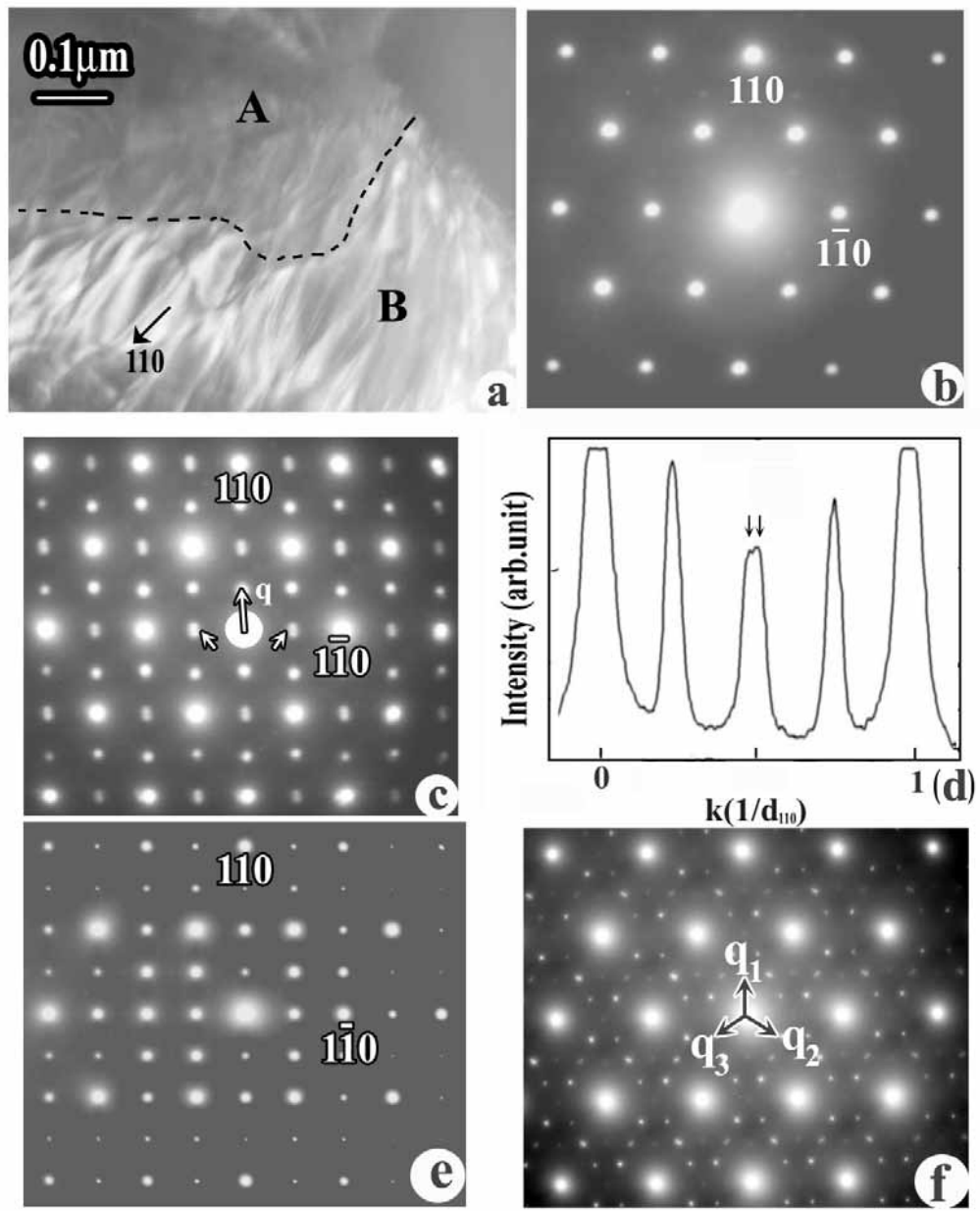

**Figure 2**

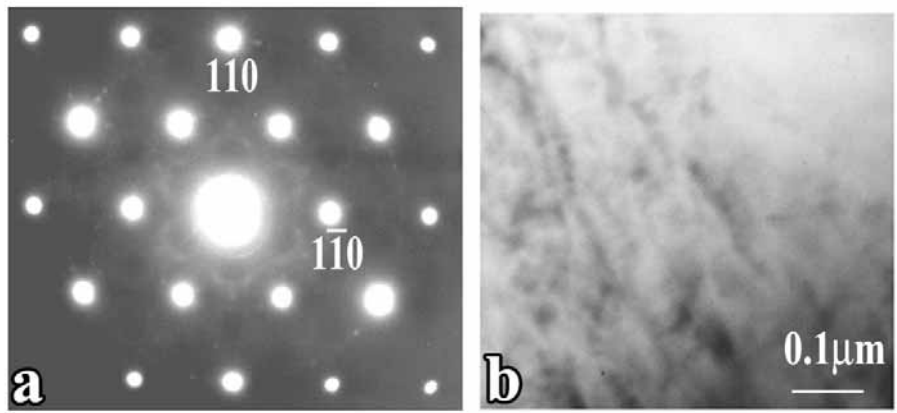

**Figure 3**

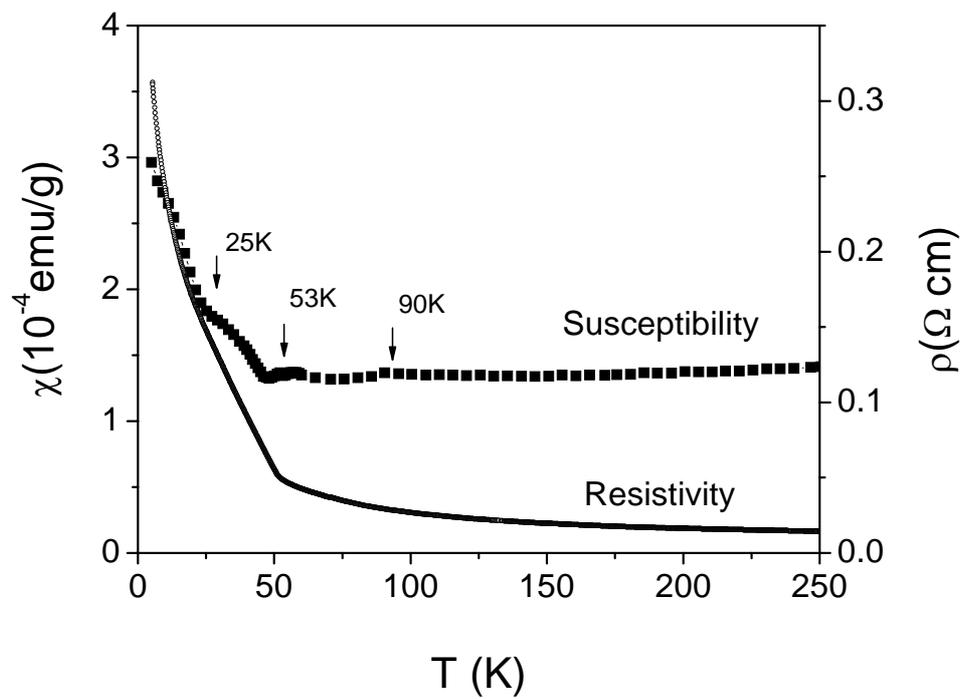

**Figure 4**

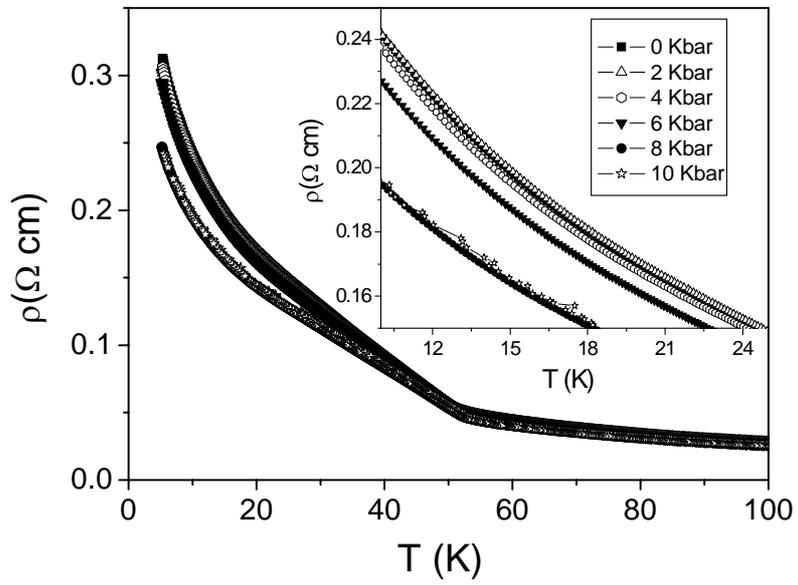

(a)

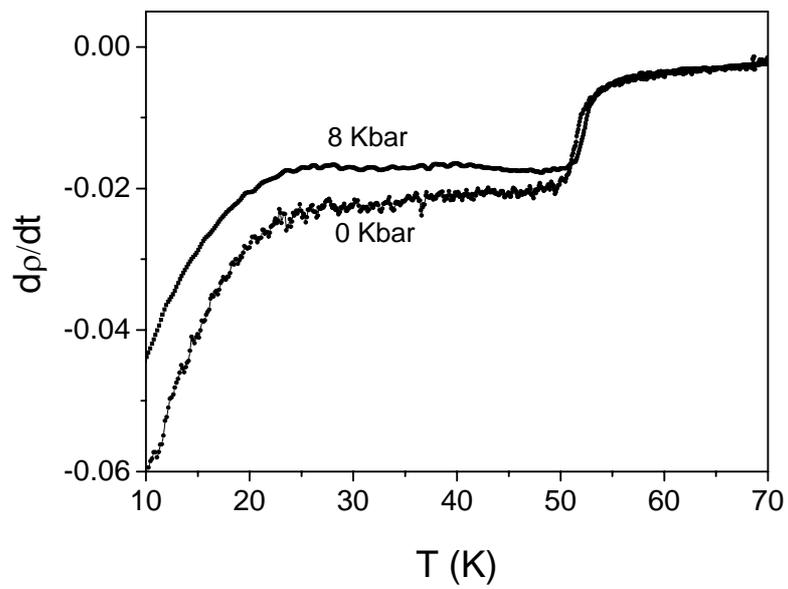

(b)

**Figure 5**

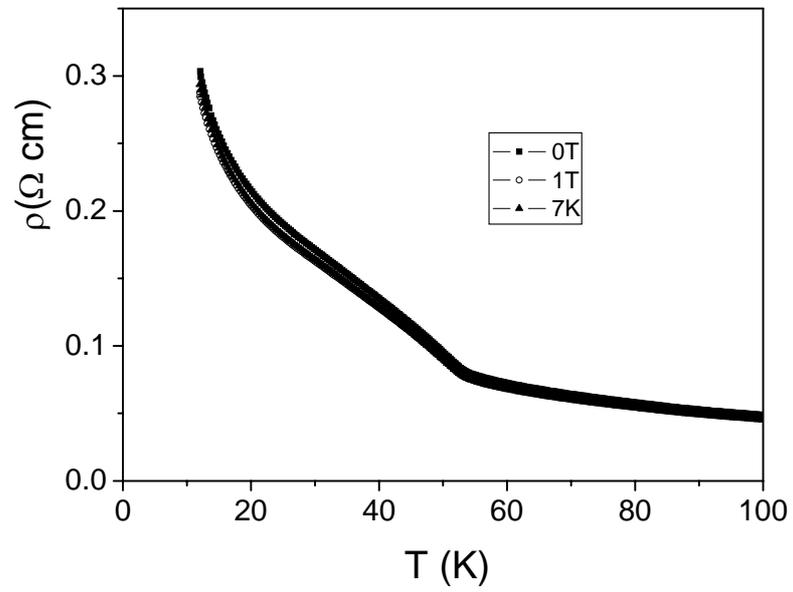

(a)

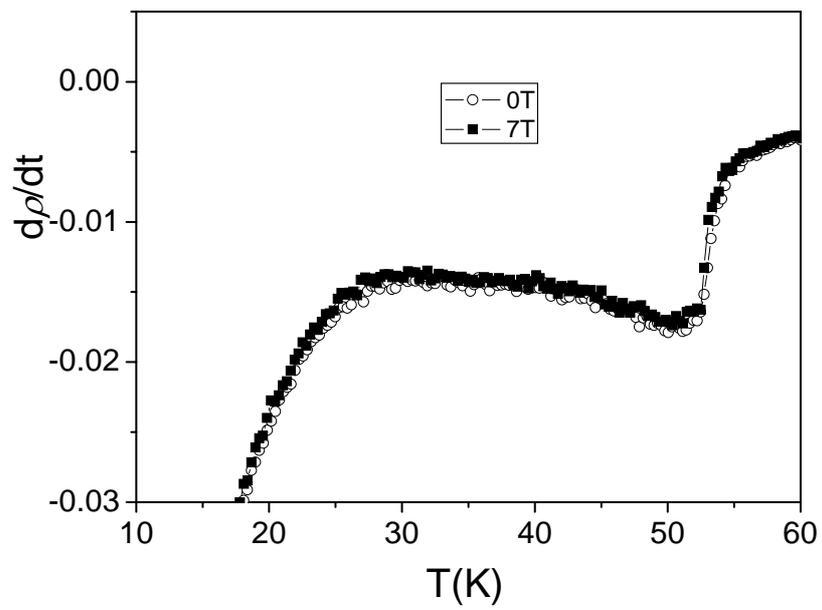

**(b)**